\documentclass{amsart}

\usepackage{graphicx}
\theoremstyle{definition}

\theoremstyle{remark}

\numberwithin{equation}{section}

\begin{document}

\title{Binary sampling ghost imaging: add random noise to fight quantization caused image quality decline}

\author{Junhui Li$^1$}
\address{$^1$ State Key Laboratory of Advanced Optical Communication Systems and Networks, School of Electronics Engineering and Computer Science, and Center for Quantum Information Technology, Peking University, Beijing 100871, China}

\author{Dongyue Yang$^2$}
\address{$^2$ School of Electronic Engineering, Beijing University of Posts and Telecommunications, Beijing 100876, China}

\author{Bin Luo$^3$}
\address{$^3$ State Key Laboratory of Information Photonics and Optical Communications, Beijing University of Posts and Telecommunications, Beijing 100876, China}

\author{Guohua Wu$^2$}

\author{Longfei Yin$^2$}

\author{Hong Guo$^1$}
\email{hongguo@pku.edu.cn}



\begin{abstract}
When the sampling data of ghost imaging is recorded with less bits, i.e., experiencing quantization, decline of image quality is observed. The less bits used, the worse image one gets. Dithering, which adds suitable random noise to the raw data before quantization, is proved to be capable of compensating image quality decline effectively, even for the extreme binary sampling case. A brief explanation and parameter optimization of dithering are given.
\end{abstract}

\maketitle

Ghost imaging (GI), contrary to ordinary spatial resolved imaging, uses the correlation between a spatial non-resolved bucket detector, and the reference arm---a non-intrusive spatial profile, either a non-contact light field \cite{Shih95}, or a modulation pattern \cite{Shapiro08}. As a price to pay, a lot of measurements are required to reconstruct a high quality image. In fact, this has become a major drawback preventing GI from practical applications, especially realtime tasks, even with the help of compressive sensing technique \cite{Bromberg09}. Considering this, sampling bit depth compression (referred as ``quantization'' hereafter for simplicity, following the terminology in \cite{Quantization98}, which should not be confused the quantization in quantum mechanics), i.e., using detectors with lower dynamic range, or recording  the measured result with less bits, would be of great significance, since the GI process can speed up when there are less data to be sampled, transported, stored, and calculated.

The extreme case of sampling bit depth compression is the binary sampling, when the output only has two values. Coincidentally, binary output is the character of single-photon avalanche diode (SPAD), one of the most sensitive device to measure light intensity. The natural combination of binary sampling and SPAD could benefit GI in both sensitivity and speed, comparing with current schemes using light intensity detectors, e.g., PIN diode and charge-coupled device (CCD). What is more, ongoing development of large scale SPAD arrays based on complementary metal-oxide-semiconductor (CMOS) fabrications offers a more compact and cost-effective alternative for the imaging applications now using intensified and electron-multiplying CCD (I-CCD \& EM-CCD) \cite{SPAD16}. Here we want to mention that, although SPADs are used in early GI experiments, e.g., \cite{Shih95}, they are not binary sampling, since each time it is the number of registered photons during a relative long period of time---known as ``time bin'', or ``detection window''---rather than the ``click or not'' two-level output, that is recorded. As for computational GI \cite{Shapiro08}, where the reference arm changed from received light field profile into modulation pattern, binary sampling would bring much convenience to the modulation, especially for the most widely used component---digital micromirror device (DMD) \cite{IEEE08}, which is in nature a two-level device and has to conduct period-multiplexing to accomplish gray scale modulation.

\begin{figure}[htbp]
\centering
\fbox{\includegraphics[width=0.9\linewidth]{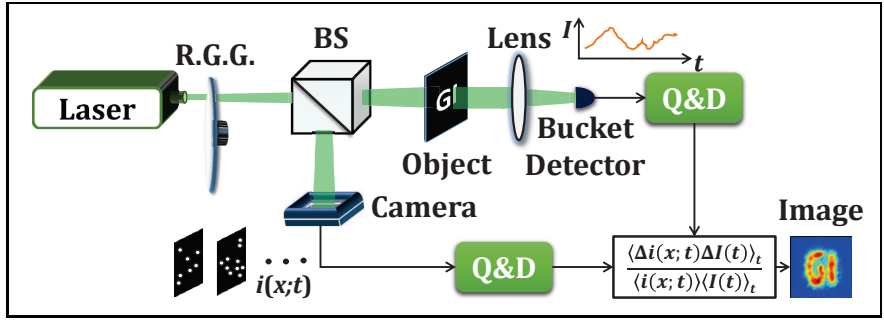}}
\caption{Experiment setup. R.G.G. = rotating ground glass. BS = beam splitter. Q\&D = quantization and dither (Fig. \ref{fig:4}(a)).}
\label{fig:1}
\end{figure}

In this Letter, we show that binary sampling GI, whose both arms only get two-level output in each measurement, is able to restore high quality images. We implement a conventional pseudo-thermal light GI setup, and apply uniform quantization \cite{Quantization98} on the recorded data of both the signal and the reference arm to achieve sampling bit depth compression (hereafter we refer the arm with object and bucket detector as ``signal arm'', and the other one without object as ``reference arm''). Quantization of the reference arm appears to do more harm on image quality than that of the signal arm. In order to compensate the image quality decline caused by quantization, additive random noise, also known as dither \cite{Dither92}, has been introduced to both arms. Dithers of different probability distribution function (PDF)---rectangular and triangular, where the latter is the summation of two independent rectangular dithers---and different amplitudes are applied to find the optimal parameter settings.

\begin{figure*}[htbp]
\centering
\fbox{\includegraphics[width=\linewidth]{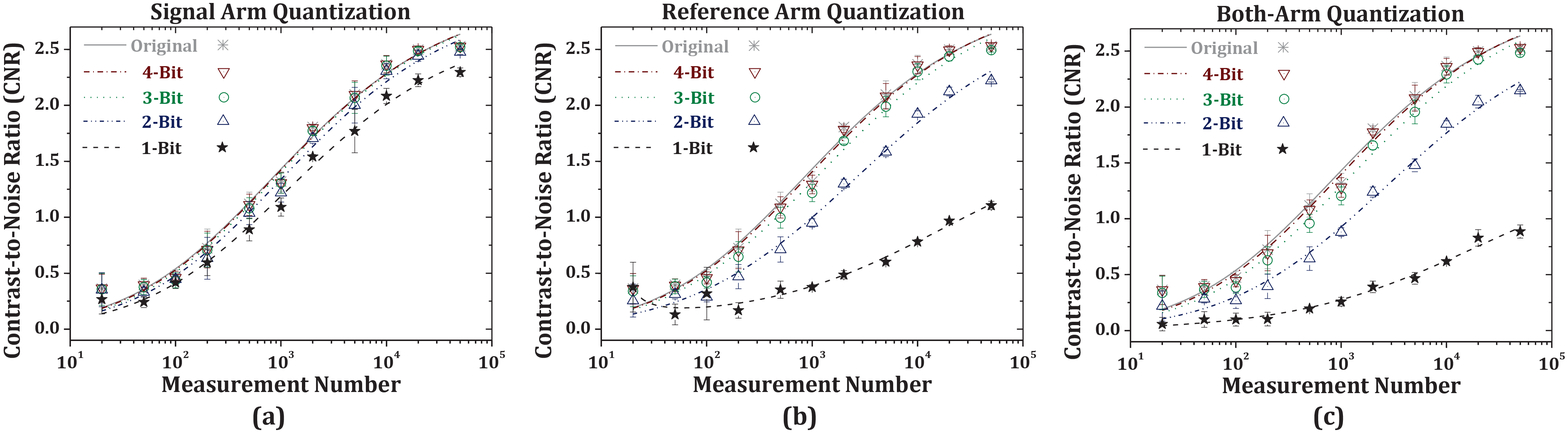}}
\caption{Reconstructed image contrast-to-noise ratio (CNR, Eq. (\ref{eq:2})) under~$n$--bit quantization in: (a) signal arm only; (b) reference arm only; (c) both arms. Original bucket signal has~$69$~values (more than~$6$~bits), while reference has~$168$~(more than~$7$~bits). The image has~$120*120$~pixels. Measured points are fitted according to \cite{arXiv}. CNR is much more sensitive to quantization on reference arm than signal arm.}
\label{fig:2}
\end{figure*}

A conventional GI setup is implemented as in Fig. \ref{fig:1}. Pseudo--thermal light, made by a 532 nm laser passing through a rotating ground glass (R.G.G.), is split into two arms by the beam splitter (BS). The signal arm penetrates a transmissive object mask, then a focus lens, to be registered as an intensity sequence~$I\left(t\right)$~by a point-like bucket detector. The spatial profile of the reference arm,~$i\left(x;t\right)$, which never touches the object, is recorded by a commercial CMOS camera synchronically with the bucket detector. The second order fluctuation correlation \cite{Shih95} between~$I\left(t\right)$~and~$i\left(x;t\right)$~yields the reconstructed image~$O\left(x\right)$,
\begin{equation}
O\left( {x} \right) \propto \frac{{{{\left\langle {\left[ {i\left( {x;t} \right) - {{\left\langle {i\left( {x;t} \right)} \right\rangle }_t}} \right] \times \left[ {I\left( t \right) - {{\left\langle {I\left( t \right)} \right\rangle }_t}} \right]} \right\rangle }_t}}}{{{{\left\langle {i\left( {x;t} \right)} \right\rangle }_t}{{\left\langle {I\left( t \right)} \right\rangle }_t}}},
\label{eq:1}
\end{equation}
where~$\left\langle  \cdot  \right\rangle_t$~denotes average over all the measurements.

We use contrast-to-noise ratio (CNR) to estimate the quality of reconstructed images. For the image~$I\left(x\right)$~of a binary object~$O\left(x\right)$, CNR reads
\begin{eqnarray}
\label{eq:2}
\rm{CNR}} = {{\left( {\sum\limits_{O\left( x \right) = 1} {I\left( x \right)}  - \sum\limits_{O\left( x \right) = 0} {I\left( x \right)} } \right)} \mathord{\left/
 {\vphantom {{\left( {\sum\limits_{O\left( x \right) = 1} {I\left( x \right)}  - \sum\limits_{O\left( x \right) = 0} {I\left( x \right)} } \right)} {\sqrt {\sigma _1^2 + \sigma _0^2} }}} \right.
 \kern-\nulldelimiterspace} {\sqrt {\sigma _1^2 + \sigma _0^2} },
\end{eqnarray}
where the numerator is the difference between the intensity summation over all the image pixels~$x$~where the object is~$1$~and~$0$, respectively, and~$\sigma_1$~and~$\sigma_0$~are the square root variance of those pixels accordingly. The higher CNR is, the better image one gets.

\begin{figure}[htbp]
\centering
\fbox{\includegraphics[width=0.95\linewidth]{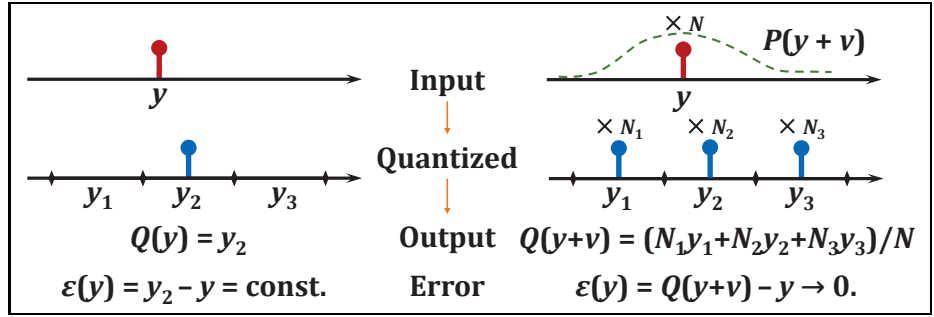}}
\caption{Left: quantizer output~$Q \left( y \right)$~is fixed for the same input~$y$, causing statistical dependent and inhomogeneous quantization error~$\varepsilon \left( y \right)$. Right: adding random dither~$\nu$~to the input randomizes the output, leading to independent and homogeneous error and thus, less distortion between input and output.}
\label{fig:3}
\end{figure}

Uniform mid-riser quantization \cite{Dither92} with the two outermost cells being semi-infinite, cf. \cite{Quantization98}, is applied to the recorded data of each measurement. The~$n$--bit quantizer~$Q$~ transfers the quasi-continuous input~$y$~into~$2^n$~equispaced discrete levels,
\begin{eqnarray}
\label{eq:3}
Q\left( y \right) = \Delta \left\lfloor {\frac{y}{\Delta }} \right\rfloor  + \frac{\Delta }{2},
\end{eqnarray}
where~$\Delta$~is the quantizer step size, and the floor operator~$\left\lfloor  \cdot \right\rfloor$~returns the greatest integer no more than its argument. We assume that~$\Delta$~is fixed, and~$\max \left( y \right) - \min \left( y \right) \le {2^n}\Delta.$~The inputs beyond this range, i.e., overloads, are set to be the outermost levels \cite{Quantization98}. In this way, our quantizer is a better model of practical detectors, which has fixed dynamic range and resolution, than those with infinite output levels appeared in previous theoretical reports, e.g., \cite{Dither92,Dither93}. We apply~$1$,~$2$,~$3$, and~$4-$bit quantization on the data of the signal arm only, the reference arm only, and both arms, respectively, and the CNR performance under varying number of measurement is shown in Fig. \ref{fig:2}. It is clear that the less bits are recorded, the worse the image quality is. An interesting finding is that compared with the signal arm, the reference arm is apparently more sensitive to the quantization, and the image quality of both arm quantization is dominantly determined by the reference arm. This may be a good news for GI in remote sensing applications \cite{Han12}, in which the reference arm can be miniaturized while the signal arm has to propagate through quite long optic path and results in an extreme weak signal. Our finding suggests that these schemes can remain a non-quantized reference arm while quantized the bucket signal even into binary output, and can still maintain high image quality.

\begin{figure}[htbp]
\centering
\fbox{\includegraphics[width=0.95\linewidth]{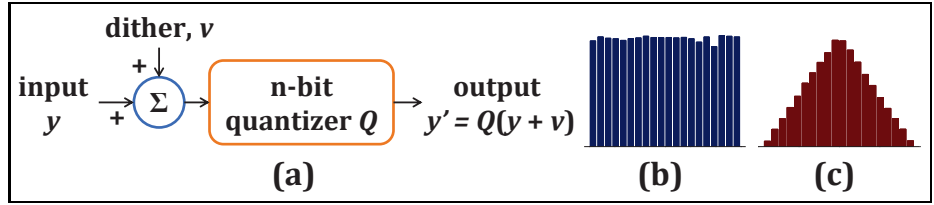}}
\caption{Dither setup: (a) Non-subtractive dithering. Quantizer~$Q$~follows Eq. (\ref{eq:3}). (b) Dither of rectangular probability distribution function (PDF). (c) Dither of triangular PDF.}
\label{fig:4}
\end{figure}

Now we turn to the extreme case where both arms are binary quantized. As Fig. \ref{fig:2} (c) suggests, image quality in this case is quite poor. It has been known that for a general quantized sampling problem, one of the major cause of the signal distortion, especially when the number of quantized levels were small, was that the quantization error~$\varepsilon$~was not statistically independent with the input signal~$y$. Introducing additive random noise, i.e., dither~$\nu$, would eliminate such dependence, therefore reduce the distortion \cite{Quantization98,Dither92,Dither93}. A brief explanation of how dither works in general is given in Fig. \ref{fig:3}. For a fixed input~$y$, quantization output~$Q \left( y \right)$~without dither will also be fixed, say,~$y_2$. The quantization error~$\varepsilon \left( y \right)$~is fully determined by the input~$y$. This strong correlation makes~$\varepsilon \left( y \right)$~inhomogeneous with respect to~$y$, results in a large distortion. When dither~$\nu$~of certain random PDF is added to the input~$y$~and turns it into a random variable~$y + \nu$, the output can have multiple values,~$y_1, y_2, y_3,$~etc. The quantization error~$\varepsilon \left( y \right)$~is thus randomized. With dithers of suitable amplitude and PDF, the average~$\varepsilon \left( y \right)$~of~$N$~samples approaches zero when~$N$~is large, regardless of specified value of the input~$y$. This randomization eliminates the statistical dependence between input~$y$~and quantization error~$\varepsilon \left( y \right)$, making the latter homogeneous again, thus reducing the distortion made by few-bit quantization~$Q \left( y \right)$~(cf. \cite{Analogdither13} ``high frequency dithering''). As for GI tasks specifically, following Eq. (\ref{eq:1}), quantization error~$\varepsilon$~is introduced to both arms after quantization, i.e.,~$i = i_0 + \varepsilon \left( i \right)$,~$I = I_0 + \varepsilon \left( I \right)$, where subscript~$0$~denotes original values before quantization. Now the numerator turns into
\begin{eqnarray}
\label{eq:4}
\begin{array}{l}
\left\langle {\left( {i - {{\left\langle i \right\rangle }_t}} \right) \cdot \left( {I - {{\left\langle I \right\rangle }_t}} \right)} \right\rangle_t  = \left\langle {\left( {{i_0} - {{\left\langle {{i_0}} \right\rangle }_t}} \right) \cdot \left( {{I_0} - {{\left\langle {{I_0}} \right\rangle }_t}} \right)} \right\rangle\\
  + {\mathop{\rm cov}} \left[ {\varepsilon \left( i \right),{I_0}} \right]
 + {\mathop{\rm cov}} \left[ {{i_0},\varepsilon \left( I \right)} \right] + {\mathop{\rm cov}} \left[ {\varepsilon \left( i \right),\varepsilon \left( I \right)} \right],
 \end{array}
\end{eqnarray}
where~${\mathop{\rm cov}} \left( {a,b} \right) = {\left\langle {ab} \right\rangle _t} - {\left\langle a \right\rangle _t}{\left\langle b \right\rangle _t}$~is the covariance of arguments~$a,b$. For the right hand side of Eq. (\ref{eq:4}), only the first term is demanding, and all the other three are errors introduced by quantization. To eliminate these errors, one has to make three pairs of quantities---$\left[ {\varepsilon \left( i \right),{I_0}} \right]$,~$\left[ {{i_0},\varepsilon \left( I \right)} \right]$, and~$\left[ {\varepsilon \left( i \right),\varepsilon \left( I \right)} \right]$~statistical independent. Noticing that~$i$~is a function of~$I$, the above requirements will be fulfilled when the quantization error~$\varepsilon \left( i \right)$~and~$\varepsilon \left( I \right)$~are independent of the inputs~$i_0$~and~$I_0$~, correspondingly. Apparently, this is a job that dither can do. That is why introducing dither can compensate image quality decline caused by quantization.

\begin{figure}[htbp]
\centering
\fbox{\includegraphics[width=0.98\linewidth]{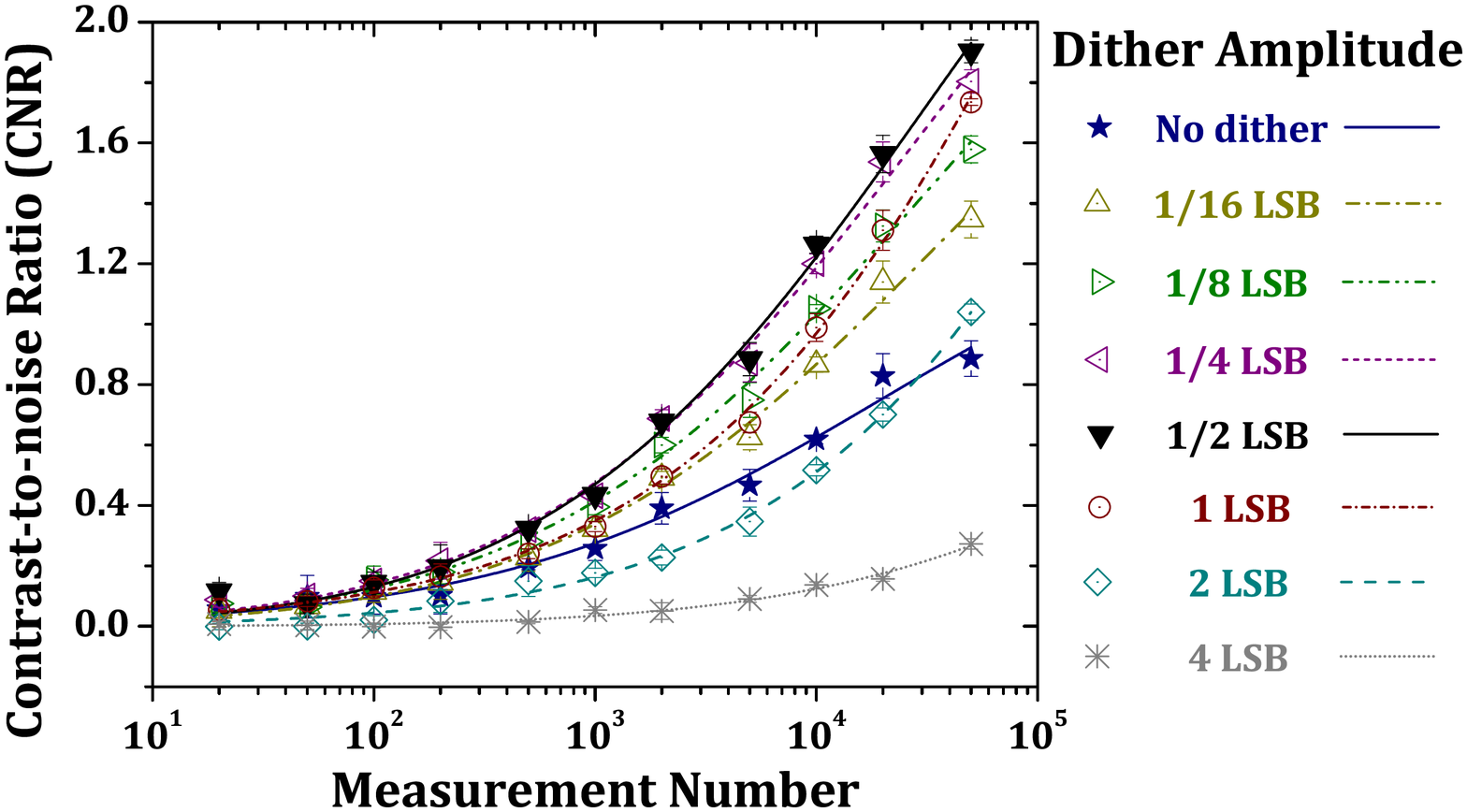}}
\caption{GI image CNR under rectangular dithering of various amplitudes. Measured points are fitted according to \cite{arXiv}.}
\label{fig:5}
\end{figure}

Now we turn to experimental verification. There are two approaches of dithering: subtractive and non-subtractive, where the former has to subtract the dither after quantization, while the latter not. Since subtractive dithering is impractical in many applications \cite{Quantization98}, we focus on non-subtractive dithering. The dithering process is shown in Fig. \ref{fig:4} (a). It is known that \cite{Dither92,Dither93}, for uniform quantization with infinite output levels, a non-subtractive dither generated by the summation of~$n$~statistically independent, zero-mean random variable uniformly distributed in a given range (i.e., of rectangular PDF), each has peak-to-peak amplitude of one least significant bit (LSB, equals to~$\Delta$~in our case), renders the first~$n$~moments of quantization error (hereafter the term ``quantization error'' follows the definition in \cite{Dither93}, rather than \cite{Dither92}),~$\varepsilon = Q\left(y+\nu\right)-y,$~independent of the input~$y$. Equivalently, the conditional mean of~$\varepsilon^n$~regarding~$y$
\begin{eqnarray}
\label{eq:5}
E\left( {{\varepsilon ^m}\left| y \right.} \right) = {\rm{const}}.,{\rm{   }}m = 1,2, \ldots ,n,
\end{eqnarray}
where~$E\left(  \cdot  \right)$~stands for expectation. Since the highest order of~$\varepsilon$~in Eq. (\ref{eq:4}) is~$2$, we only have to deal with~$E\left( {{\varepsilon}\left| y \right.} \right)$~and~$E\left( {{\varepsilon^2}\left| y \right.} \right)$~. Corresponding dithering has two types only: rectangular PDF, and the summation of two independent rectangular ones, whose PDF is triangular---convolution of two rectangular PDFs, as Fig. \ref{fig:4} (b) and (c) shows, respectively. CNR of GI image under varying measurement number when both arms binary quantized, applying rectangular PDF dithering of different peak-to-peak amplitudes are summarized in Fig. \ref{fig:5}. The result of triangular dithering is similar. There is a range of dither amplitudes when the image quality is improved comparing with the original non-dithering situation. The improvement is even larger when the number of measurement is large. Noticing that the right most point corresponds to measurement number about four times of the number of image pixels, it suggests our dithering method is even more advantageous in the over-sampling scenario. Meanwhile, CNR drops quickly, to even worse than the undithered case when the amplitudes goes beyond that range. This hints the importance to control the dither amplitude.

\begin{figure}[htbp]
\centering
\fbox{\includegraphics[width=0.8\linewidth]{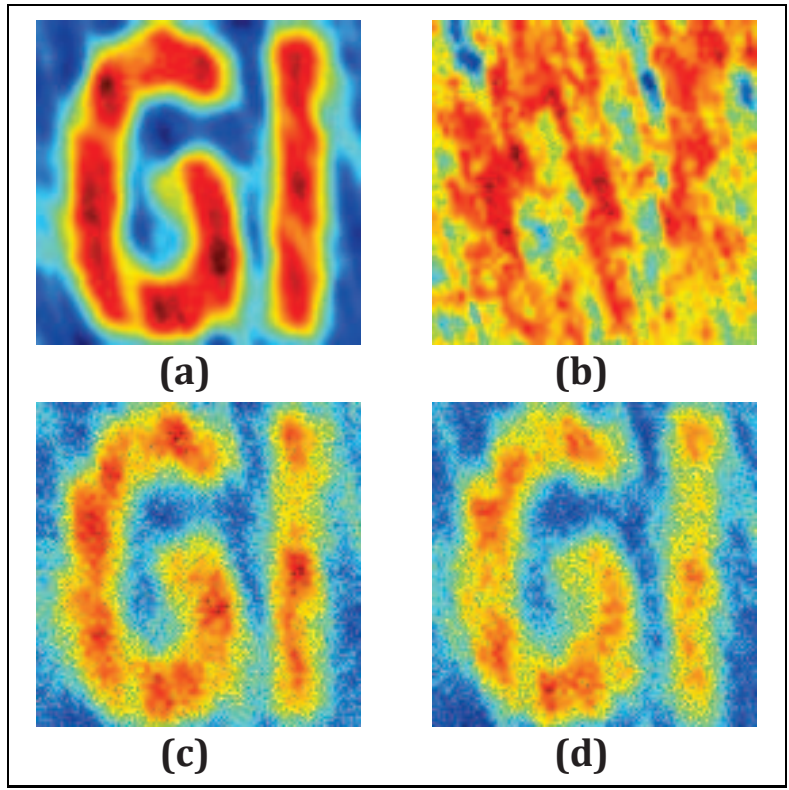}}
\caption{Optimum reconstructed images with both arms: (a) not quantized; (b) binary quantized and not dithered; (c) binary quantized and rectangular dithered; (d) binary quantized and triangular dithered. All the images are normalized so that summations over all the pixels equal to unity.}
\label{fig:6}
\end{figure}
\begin{figure*}[htbp]
\centering
\fbox{\includegraphics[width=\linewidth]{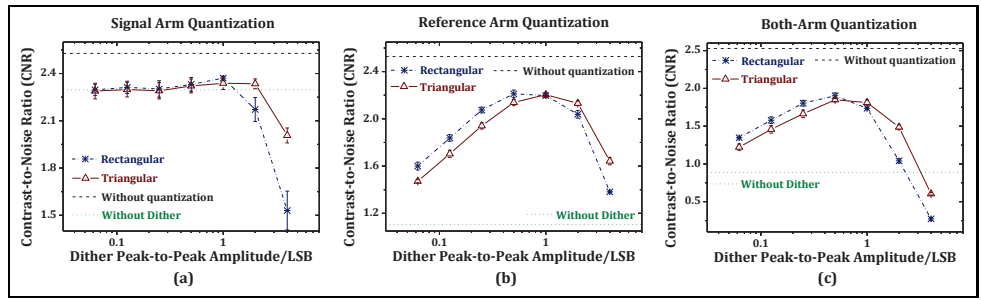}}
\caption{Peak CNR under various dithering amplitudes with: (a) only signal arm; (b) only reference arm; (c) both arms binary quantized. Performance of rectangular and triangular dithering are similar. Optimum dithering amplitude is around~$0.5 - 1$~LSB.}
\label{fig:7}
\end{figure*}

The best reconstructed image with both arms binary quantized under suitable dithering are shown in Fig. \ref{fig:6}, comparing with the non-quantized image and the undithered one. Apparently, dithered images are much better than the undithered one, indicating that high quality, binary sampling GI is quite promising. The maximum CNR under different quantization situations and various dithering amplitudes are summarized in Fig. \ref{fig:7}. Other than nuisance difference, rectangular and triangular dithering have nearly the same performance, suggesting that the first order terms of quantization error, rather than the second order terms, dominates Eq. (\ref{eq:4}). The  optimum dither amplitude of both dithering is between~0.5~and~1~LSB.

Our findings above would be important for GI in several ways. When the received light is very weak, highly sophisticated apparatus capable of photon number resolving are implemented currently, such as photomultiplier tube (PMT) and I-CCD/EM-CCD. Since under suitable dither, binary sampling GI can restore high quality image as we showed, the current setup can be replaced by a much more compact and cost-effective scheme based on SPAD and its array. On the other hand, sampling bit depth compression increases GI's robustness against possible noise, since the quantized output is immune to a fluctuation whose amplitude is much smaller than~$1$~LSB. Besides, reduction of total data amount to be collected, stored, transported and calculated after quantization may be vital to GI speed improvement. As for computational GI, DMD---the most widely used modulation component---has to await several periods every time to realize modulation with gray scales. A binary modulation protocol rooted in our binary sampling scheme would eliminate the necessity of period-multiplexing, and thus improve the speed of modulation.

In conclusion, we show that GI would suffer image quality decline after quantization, and dithering could restore high quality image even both arms are binary quantized. Besides, for a conventional GI setup, the reference arm is more sensitive to quantization than the signal arm. Optimum dithering settings are found. Our findings pave the way for applying binary sampling in various GI applications, and may benefit in many ways.

\subsection*{Funding}
National Natural Science Foundation of China (61631014, 61401036, 61471051, 61531003); National Science Fund for Distinguished Young Scholars of China (61225003); China Postdoctoral Science Foundation (2015M580008); Youth Research and Innovation Program of BUPT (2015RC12); PhD Students' Overseas Research Program of Peking University, China.

\bibliographystyle{amsplain}

\end{document}